\documentclass[onecolumn,showpacs,preprintnumbers,amssymb]{revtex4-1}
\usepackage{graphicx,amsmath}
\usepackage{dcolumn}
\usepackage{bm}
\usepackage{amsmath}

\begin{document}


\title{Singlet S-wave superfluidity of proton in neutron star matter}


\author{Xu Yan$^{1**}$,\quad Zhang~Xiao-Jun$^{1}$,\quad Fan~Cun-Bo$^{1}$,\quad Tmurbagan Bao$^{2}$,\quad Huang~Xiu-Lin$^{1**}$,\quad Liu~Cheng-Zhi$^{1**}$}

\affiliation{%
1 Changchun Observatory, National Astronomical Observatories, Chinese Academy of Sciences, Changchun 130117, China \\
2 College of Physics and Electronic Information,Inner Mongolia University for the Nationalities, Tongliao 028043, China}%
\date{\today}


\date{\today}

\begin{abstract}
The possible $^{1}S_{0}$ protonic superfluidity is investigated in
neutron star matter, and the corresponding energy gap as a function
of baryonic density is calculated on the basis of BCS gap equation.
We have discussed particularly the influence of hyperon degrees of
freedom on $^{1}S_{0}$ protonic superfluidity. It is found that the
appearance of hyperons leads to a slight decrease of $^{1}S_{0}$
protonic pairing energy gap in most density range of existing
$^{1}S_{0}$ protonic superfluidity. However, when the baryonic
density $\rho_{B}>$0.377 (or 0.409) fm${^{-3}}$ for TM1 (or TMA)
parameter set, $^{1}S_{0}$ protonic pairing energy gap is
significantly larger than the corresponding values without hyperons.
And the baryonic density range of existing $^{1}S_{0}$ protonic
superfluidity is widen due to the appearance of hyperons. In our
results, the hyperons not only change the EOS and bulk properties
but also change the size and baryon density range of $^{1}S_{0}$
protonic superfluidity in neutron star matter.
\end{abstract}


\maketitle


\section{Introducton}
The dense neutron star(NS) matter shows us an interesting subject to
study the properties of nucleon matter with the density higher than
nuclear saturation density $\rho_{0}$. There, it has pointed that
the various new degrees of freedom, such as hyperons, quarks and
their mixed phases, realize according to the
density\cite{Meng2006,Bednarek2005,Yang2008}. It is well known that
NS matter has the properties of the strong degeneracy and exists the
attractive interaction between two baryons, which are the conditions
for the occurrence of superfluid states in Fermi systems. Thus, NS
is already considered as the key laboratories of various
superfluidity in nuclear
matter\cite{Amundsen1985,Chen1993,ZhaoEnGuang2011,
Zuo04,Zuo2010,XuFR2012,Tanigawa2004,Lombardo2005,Lombardo2006,xuyancpl20131}.
In recent years, many studies have been focussing on $^{1}S_{0}$
protonic superfluidity. Since the protonic superfluid states as well
as their pairing strength can greatly suppress the neutrino
processes involving nucleons about $10^{5}-10^{6}$ years of NS
cooling phases, affect the properties of rotating dynamics,
post-glitch timing observations and possible vertex pinning of
NSs\cite{Gnedin1993,ZhengXiaoPing2006,ZhengXiaoPing2010,ZhouXia2013,chenwei2006,Yakovlev3,xuyandlzt,xuyancpl20132,xuyancsb2014}.

The possibility of protonic superfluid states in NS matter is first
suggested by Migdal in 1960\cite{Migdal1960}. The interaction
between two protons is the combination of strong repulsive
short-range interaction and weaker attractive long-range
interaction. In proton matter, when the interparticle distance is
much larger than the range of the repulsive interaction, protons
will condense into superfluid states due to the attractive
interaction. For the size of $^{1}S_{0}$ protonic pairing energy
gap, the calculations based on the microscopic theory have already
carried out a great amount of work. All the numerical calculations
yield qualitatively similar ranges for the appearance of $^{1}S_{0}$
protonic pairing. However, obtaining the exact numerical results for
$^{1}S_{0}$ protonic pairing energy gap and evaluating its
quantitative influence on NS have been proved to be a hard problem,
which is due to that there are many uncertainties about the
proton-proton(pp) interactions in NS matter, methods of
approximation, paucity of experimental data in extreme conditions
and so on. During this period, lots of researches of NS have been
performed adopting various frameworks. Currently, many relativistic
models call attention in researches on NS since they are well suited
to describe NS in accord with the special relativity. The most
commonly used among them is the relativistic mean field(RMF) model,
extremely successful in nuclear matter
studies\cite{Glendenning1985,Glendenning1991,shenH2010}.

This article mainly does the following work. We study NS matter for
two cases: (i)NS is made up of neutron, proton, electron and muon
only (npe$\mu$), (ii)NS is composed of nucleons, hyperons($\Lambda$
and $\Xi$), electron and muon (npHe$\mu$). Our model excludes
$\Sigma$ hyperons on account of the remaining uncertainty of the
form of $\Sigma$ potential in nuclear matter at the saturation
density\cite{Batty,Simon}. We use the RMF theory to describe the
properties of NS. The $^{1}S_{0}$ protonic pairing energy gap is
calculated by the Reid soft core(RSC)
potential\cite{Nishizaki1991,Wambach1993}. We mainly focus on the
influence of hyperon degrees of freedom on $^{1}S_{0}$ protonic
pairing energy gap in NS matter.

\section{The models}
In the RMF theory, baryonic interactions are described by the
exchanged mesons including isoscalar scalar and vector mesons
$\sigma$ and $\omega$, an isovector vector meson $\rho$, two
additional strange mesons $\sigma ^{*}$ and $\phi$. The total
lagrangian density of NS matter
is\cite{Glendenning1985,Glendenning1991,shenH2010},
\begin{multline}
\hspace{-5mm}
L=\sum_B\overline{\psi}_B[i\gamma_\mu\partial^\mu-(m_B-g_{\sigma
B}\sigma-g_{\sigma^*B\sigma^*}) -g_{\rho
B}\gamma_\mu{\boldsymbol{\tau}}\cdot{\boldsymbol{\rho}}^\mu\\
-g_{\omega B}\gamma_\mu\omega^\mu-g_{\phi B}\gamma_\mu\phi^\mu
 ]\psi_B
 +\frac{1}{2}(\partial_\mu\sigma\partial^\mu\sigma-m_\sigma^2\sigma^2)-\frac{1}{3}a \sigma
 ^{3}\\
 -\frac{1}{4}b \sigma^4
   +\frac{1}{2}m_\omega^2
  \omega_\mu\omega^\mu+\frac{1}{4}c_3(\omega_\mu\omega^\mu)^2+
\frac{1}{2}m_\rho^2{\boldsymbol{\rho}}_\mu{\boldsymbol{\rho}}^\mu
-\frac{1}{4}F^{\mu v}F_{\mu v}\\-\frac{1}{4}G^{\mu v}G_{\mu v}+
+\frac{1}{2}(\partial_v\sigma^*\partial^v\sigma^*-m^2_{\sigma^*}\sigma^{*2})
-\frac{1}{4}S^{\mu v}S_{\mu
v}\\+\frac{1}{2}m^2_{\phi}\phi_\mu\phi^\mu+\sum_l\overline{\psi}_l[i\gamma_\mu\partial^\mu-m_l]\psi_l.
\end{multline}
Here the field tensors of the vector mesons, $\sigma$ and $\omega$,
are denoted by $F_{\mu
v}=\partial_\mu\omega_v-\partial_v\omega_\mu$, and $G_{\mu
v}=\partial_\mu{\boldsymbol{\rho}}_v-\partial_v{\boldsymbol{\rho}}_\mu$.

The meson fields are seen as classical fields and field operators
are instead of their expectation values in the RMF approximation.
The field equations derived from the Lagrange function are
\begin{eqnarray}
\hspace{-2mm}
\sum_B g_{\sigma B}\rho_{SB}=m_\sigma^2\sigma+a\sigma^2+b\sigma^3,\\
\sum_B g_{\omega B}\rho_B=m_\omega^2\omega_0+c_{3}\omega^{3}_{0},\\
\sum_B g_{\rho B}\rho_{B}I_{3B} =m_{\rho}^2\rho_0\\
\sum_B g_{\sigma^* B}\rho_{SB}=m_{\sigma^*}^2\sigma^*,\\
\sum_B g_{\phi B}\rho_{B}=m_\phi^2\phi_0.
\end{eqnarray}
Here $I_{3B}$ denotes baryonic isospin projection. $\rho_{SB}$ and
$\rho_{B}$ are baryonic scalar and vector densities, respectively.
They have the following form,
\begin{eqnarray}
\rho_{SB}=\frac{2J_{B}+1}{2\pi^{2}}\int_0^{k_{FB}}\frac{m_{B}^{*}}{\sqrt{k^{2}+m_{B}^{*2}}}k^{2}dk,
\nonumber
\\
\rho_{B}=\frac{k_{FB}^{3}}{3\pi^2}.
\end{eqnarray}
Here $J_{B}$ is baryonic spin projection, $k_{FB}$ is baryonic Fermi
momentum,
$m_B^*=m_{B}-g_{\sigma_{B}}\sigma-g_{\sigma_{B}^{*}}\sigma^{*}$ is
baryonic effective mass.

For NS matter consisting of n, p, $\Lambda$, $\Xi^{0}$, $\Xi^{-}$,
e, $\mu$, the charge neutrality condition is given by
\begin{eqnarray}
\rho_{p}=\rho_{\Xi^{-}}+\rho_{e}+\rho_{\mu}
\end{eqnarray}
The $\beta$ equilibrium conditions are expressed as
\begin{eqnarray}
\mu_{p}=\mu_{n}-\mu_{e}, ~~~\mu_{\Xi^{-}}=\mu_{n}+\mu_{e} \nonumber
\\
\mu_{\Lambda}=\mu_{\Xi^{0}}=\mu_{n}, ~~~\mu_{\mu}=\mu_{e}.
\end{eqnarray}

At zero temperature the chemical potential of baryon and lepton are
written by
\begin{eqnarray}
\mu_{B}=\sqrt{k_{FB}^2+{m_B^*}^2}+g_{\omega B} \omega_{0}+g_{\rho B}
\rho_{03} I_{3B}+ g_{\phi B}\phi_{0} \nonumber,
\\
\mu_{l}=\sqrt{k_{Fl}^2+{m_l}^2}. \quad
\end{eqnarray}

The $^{1}S_{0}$ protonic pairing energy gap $\Delta_{p}$ can be
obtained by solving the BCS gap equation,
\begin{eqnarray}
\hspace{-2mm}
\Delta_{p}(k)=-\frac{1}{4\pi^{2}}\int{k^{'2}dk^{'}\frac{V(k,k^{'})\Delta_{p}(k^{'})}{\sqrt{\varepsilon^{2}(k^{'})+\Delta_{p}^{2}(k^{'})}}},
\end{eqnarray}
where $\varepsilon(k^{'})=E_{p}(k^{'})-E_{p}(k_{Fp}^{'})$,
$E_{p}(k^{'})=\sqrt{{k^{'}}^2 + {m_p^{*}}^2}+g_{\omega p} \omega_{0}+g_{\rho p} \rho_{03} I_{3p}+g_{\phi p} \phi_{0}$ is protonic single particle energy. For the pp interaction, we adopt
the RSC potential which is well suited for applying in NS matter.
$V(k,k^{'})$ is defined the matrix element of $^{1}S_{0}$ component
of RSC potential in momentum space,

\begin{equation}\label{2}
  V(k,k^{'})=\langle{k}|V(^{1}S_{0})|{k^{'}}\rangle=4\pi\int{r^{2}drj_{0}(kr)V_{pp}(r)j_{0}(k^{'}r)}.
\end{equation}

Here $V_{pp}(r)$ is  $^{1}S_{0}$ pp interaction potential in
coordinate space. It is expressed in the five-range Gaussian and
depends on $\rho_{B}$, asymmetry parameter
$\alpha=(\rho_{n}-\rho_{p})/\rho_{N}$ and two-nucleon state $\beta$
and $\gamma$ as discussed in Refs\cite{Nishizaki1991,Wambach1993},
\begin{eqnarray}
\hspace{-2mm}
V_{NN}(r)=\sum_{i=1}^{5}c_{i}(\rho_{N},\alpha,\beta,\gamma)e^{\frac{-r^{2}}{\lambda_{i}^{2}}}.
\end{eqnarray}

The critical temperature $T_{cp}$ of $^{1}S_{0}$ protonic
superfluidity is given by its energy gap $\Delta_{p}$ at zero
temperature approximation\cite{Takatsuka2004},
\begin{equation}
T_{cp}\doteq 0.66 \Delta_{p}.
\end{equation}

Combining Eqs.(1)-(6) with the charge neutrality and $\beta$
equilibrium conditions, Eqs.(8) and (9), we can solve the system
with a fixed $\rho_{B}$. Thus,
$^{1}S_{0}$ protonic pairing energy
gap $\Delta_{p}$ and critical temperature $T_{cp}$ can be obtained
from Eqs.(11)-(14).

\section{Discussion}

As mentioned above, the onset of $^{1}S_{0}$ protonic superfluidity
is determined by the energy gap function $\Delta_{p}$. Theoretical
calculation of $\Delta_{p}$ is sensitively dependent on the model of
pp interaction and many-body theory adopted. In this paper, we use
the RSC potential as an example to analyze $^{1}S_{0}$ protonic
superfluidity in npe$\mu$ and npHe$\mu$ matter, respectively. We
mainly focus on the influence of hyperon degrees of freedom on
$^{1}S_{0}$ protonic superfluidity. In order to make the results
more clearly, we employ two successful parameter sets of TM1 and TMA
to calculate separately $^{1}S_{0}$ protonic pairing energy gap in
NS matter. The parameter sets are listed in Table I and II. For the
vector couplings of hyperons, we take the relations derived from
SU(6) quark model(see Ref\cite{Bednarek2005,Yang2008,xuyancpl2012}
for details),
\begin{align*}
\frac{2}{3}g_{\omega N}=g_{\omega\Lambda}=
2g_{\omega\Xi},\hspace{1mm} g_{\rho N}= g_{\rho\Xi},\\
2g_{\phi\Lambda}=g_{\phi\Xi}=-\frac{2\sqrt{2}}{3}g_{\omega N}.
\end{align*}

\begin{table}
\caption{\label{tab:Table I}The coupling constants of the
meson-nucleon of TM1 and TMA sets, the masses are unit of
MeV\cite{Bednarek2005,Yang2008}. }
\begin{ruledtabular}
\begin{tabular}{llllllll}
Set&$m_{\sigma}$&$g_{\sigma N}$&$g_{\omega N}$&$g_{\rho N}$&a&b&$c_3$\\
\hline
TM1&511.198&10.0289&12.6139&4.6322&7.2325&0.6183&71.3075\\
TMA&519.151&10.055&12.842&3.8&0.328&38.862&151.59\\
\end{tabular}
\end{ruledtabular}
\end{table}

\begin{table}
\caption{\label{tab:Table II} The scalar coupling constants for
hyperons in two sets, the masses are unit of
MeV\cite{Bednarek2005,Yang2008}. }
\begin{ruledtabular}
\begin{tabular}{lllllll}
Set&$m_{\omega}$&$m_{\rho}$&$g_{\sigma \Lambda}$&$g_{\sigma\Xi}$&$g_{\sigma^{*}\Lambda}$&$g_{\sigma^{*}\Xi}$\\
\hline
TM1&783.0&770.0&6.2380&3.1992&3.7257&11.5092\\
TMA&781.95&768.1&6.2421&3.2075&4.3276&11.7314\\
\end{tabular}
\end{ruledtabular}
\end{table}

\begin{figure}
\includegraphics[width=0.43\textwidth]{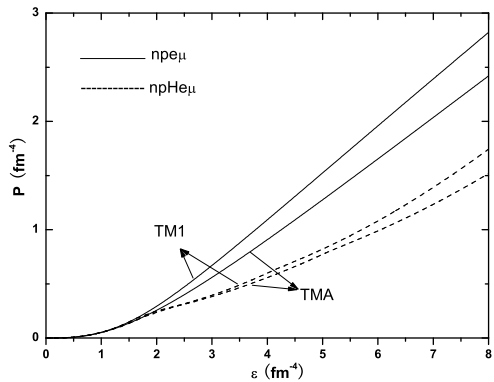}
\caption{The EOS in npe$\mu$ and npHe$\mu$ matter for TM1 and TMA
parameter sets.} \label{fig:1}
\end{figure}

\begin{figure}
\includegraphics[width=0.43\textwidth]{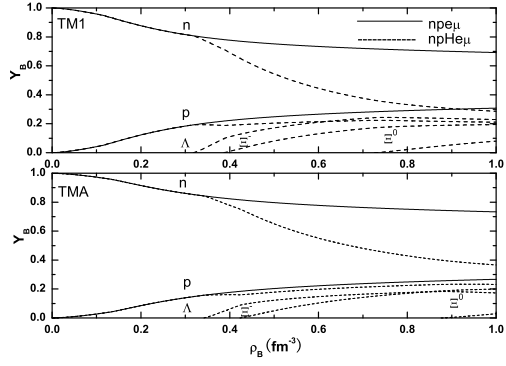}
\caption{Baryonic fraction $Y_{i}=\rho_{i}/\rho_{B}$ as a function
of baryonic density $\rho_{B}$ in npe$\mu$ and npHe$\mu$ matter for
TM1(top panel) and TMA (bottom panel) parameter sets .}
\label{fig:2}
\end{figure}

\begin{figure}
\includegraphics[width=0.43\textwidth]{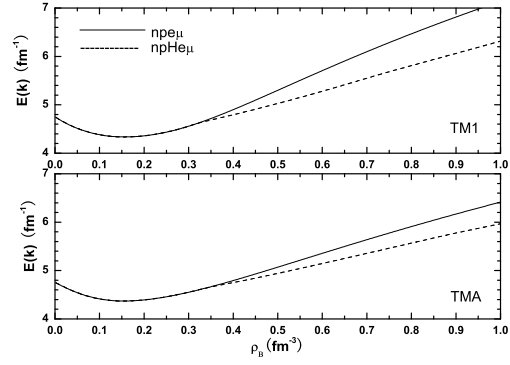}
\caption{Protonic single-particle energy $E_{p}(k)$ vs baryonic
density $\rho_{B}$ in npe$\mu$ and npHe$\mu$ matter for TM1(top
panel) and TMA (bottom panel) parameter sets.} \label{fig:3}
\end{figure}

\begin{figure}
\includegraphics[width=0.43\textwidth]{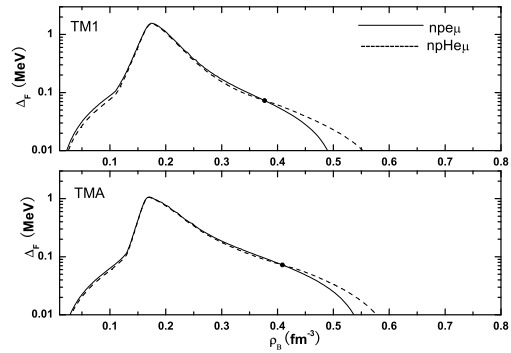}
\caption{$^{1}S_{0}$ protonic pairing energy gap $\Delta_{p}(k)$ at
the Fermi surface as a function of baryonic density $\rho_{B}$ in
npe$\mu$ and npHe$\mu$ matter for TM1(top panel) and TMA (bottom
panel) parameter sets.} \label{fig:4}
\end{figure}

Fig.1 shows the EOS, pressure $P$ versus energy density
$\varepsilon$ in npe$\mu$ and npHe$\mu$ matter, for parameter sets
TM1 and TMA. From Fig.1 one can see that at low densities the EOSs
are unchanged in npe$\mu$ and npHe$\mu$ matter for two parameter
sets. However, as baryonic density increases, the $\Lambda$,
$\Xi^{-}$, $\Xi^{0}$ appear one by one. Along with it, the EOS in
npHe$\mu$ matter gets softer than the EOS in npe$\mu$ matter for the
two parameter sets. And TMA set makes the EOS more softer in
npe$\mu$ and npHe$\mu$ matter. The soft EOSs must cause significant
changes of the bulk properties of NS matter, which must change the
the size and baryonic density range of $^{1}S_{0}$ protonic pairing
energy gap. It can be seen in Eq.(11), the protonic Fermi momentum
and single particle energy play the vitally important role in
$^{1}S_{0}$ protonic pairing energy gap. The Fig.2 represents the
results of self-consistent calculation of baryonic particle fraction
as a function of baryonic density $\rho_B$ in npe$\mu$ and npHe$\mu$
matter, for parameter sets TM1 and TMA. It can be see from Fig.2
that the change of EOS(see Fig.1) changes protonic fraction in NS
matter, that is, the appearance of hyperons makes protonic fraction
$Y_{p}$ decrease for two parameter sets. This is because that the
occurrence of hyperons suppresses protonic fraction due to the
conditions of the charge neutrality and $\beta$ equilibrium(see
Eqs.(8) and (9)) in NS matter. Then according to Eq.(7), when
hyperons appear in NS, $k_{Fp}$ becomes smaller than the
corresponding values in npe$\mu$ matter. Fig.3 shows protonic single
particle energy $E_{p}$ as a function of baryonic density $\rho_B$
in npe$\mu$ and npHe$\mu$ matter, for parameter sets TM1 and TMA. In
Fig.3, one can see that $E_{p}$ in npHe$\mu$ matter is also less the
corresponding values in npe$\mu$ matter for two parameter sets. This
is due to the decrease of protonic Fermi momentum in npHe$\mu$
matter(see Fig.2).

\begin{table}
\caption{\label{tab:Table III}The peak values of $^{1}S_{0}$
protonic pairing energy gap $\Delta_{p}^{max}$ and the corresponding
critical temperature $T_{cp}^{max}$, baryonic density $\rho_{B}$ in
npe$\mu$ and npHe$\mu$ matter, for TM1 and TMA parameter sets.}
\begin{ruledtabular}
\begin{tabular}{llll}
Set&$\Delta_{p}^{max}$&$T_{cp}^{max}$&$\rho_{B}$\\
\hline
TM1 npe$\mu$&1.560&$1.030\times10^{10}$&0.174\\
TM1 npHe$\mu$&1.528&$1.008\times10^{10}$&0.175\\
TMA npe$\mu$&1.068&$7.049\times10^{9}$&0.169\\
TMA npHe$\mu$&1.052&$6.943\times10^{9}$&0.170\\
\end{tabular}
\end{ruledtabular}
\end{table}

Up to now, we do not known $^{1}S_{0}$ protonic pairing energy gap
increase or decrease, if hyperons appear in NS matter. As the
uncertainty of the pp interaction, we calculate $\Delta_{p}$ on the
basis of the RSC potential. We concentrate on the influence of
hyperon degrees of freedom on $^{1}S_{0}$ protonic pairing energy
gap in NS matter. The peak values of $^{1}S_{0}$ protonic pairing
energy gap $\Delta_{p}^{max}$ and the corresponding critical
temperature $T_{cp}^{max}$, baryonic density $\rho_{B}$ in npe$\mu$
and npHe$\mu$ matter are listed in Table III, for the TM1 and TMA
parameter sets. As shown in Table III, the appearance of hyperons
makes $\Delta_{p}^{max}$ and $T_{cp}^{max}$ decrease which will
inevitably cause the cooling rate of NS changing. In Fig.4, we show
$^{1}S_{0}$ protonic pairing energy gap $\Delta_{p}$ as a function
of baryonic density $\rho_B$ in npe$\mu$ and npHe$\mu$ matter, for
parameter sets TM1 and TMA. As seen from Fig.4, the protonic pairing
energy gap $\Delta_{p}$ as the function of $\rho_B$ is a typical
bell-shaped curve from zero to a maximum value to zero again. The
change of $\Delta_{p}$ in behavior is due to the reduction of mean
interparticle distance. From Fig.4, we also can see that $^{1}S_{0}$
protonic superfluidity appears within the baryonic density range of
$\rho_{B}$ = 0.0 -- 0.649 (0.685) fm$^{-3}$ and $\rho_{B}$ = 0.0 --
0.953 (0.853) fm$^{-3}$ for parameter sets TM1( or TMA) in npe$\mu$
and npHe$\mu$ matter, respectively. It is clear from these data that
the appearance of hyperons makes baryonic density range of
$^{1}S_{0}$ protonic superfluidity widen. Such baryonic density
ranges can cover or partially cover the cores of NSs and are highly
relevant to the direct Urca processes involving nucleons which play
a leading role in NS cooling. In particular, from Fig.4, the
appearance of hyperons also leads to $^{1}S_{0}$ protonic pairing
energy gap in npHe$\mu$ matter obviously larger than the
corresponding values in npe$\mu$ matter when baryonic density
$\rho_{B}\geq$ 0.377 (0.409) fm$^{-3}$ for TM1(or TMA) parameter set
(see the two dots shown in Fig.4). According to Eq.(14), the
critical temperature of $^{1}S_{0}$ protonic superfluidity
positively increases which could further suppress the cooling rate
of NS.

\section{Conclusion}
We study the influence of hyperon degrees of freedom on the size and
baryonic density range of $^{1}S_{0}$ protonic paring energy gap by
adopting the RMF and BCS theories in NS matter. It is shown that the
appearance of hyperons makes the peak values of $^{1}S_{0}$ protonic
pairing energy gap and critical temperature decrease. And baryonic
density range of existing $^{1}S_{0}$ protonic superfluidity is
widen from 0.0 -- 0.649 (0.685) fm$^{-3}$ in npe$\mu$ matter to 0.0
-- 0.953(0.853)fm$^{-3}$ for TM1 (or TMA) parameter sets in
npHe$\mu$ matter. In addition, when the baryonic density
$\rho_{B}\geq$ 0.377 (0.409) fm$^{-3}$ for TM1(or TMA) parameter
set, the appearance of hyperons leads to $^{1}S_{0}$ protonic
pairing energy gap obviously larger than the corresponding values in
npe$\mu$ matter which could further suppress the cooling rate of
NSs. In our results, the appearance of hyperons in NS matter not
only changes the EOS and bulk properties but also changes the
properties of $^{1}S_{0}$ protonic superfluidity.

\end{document}